\documentclass[11pt,a4paper,reqno]{amsart}
\usepackage{graphicx}
\usepackage{epsfig}
\usepackage{amsmath}
\usepackage{amsfonts}
\usepackage{amssymb}
\usepackage{amscd}

\begin{document}

\begin{minipage}[b]{0.5\linewidth}
{\includegraphics[height=0.82in,width=5.94in]{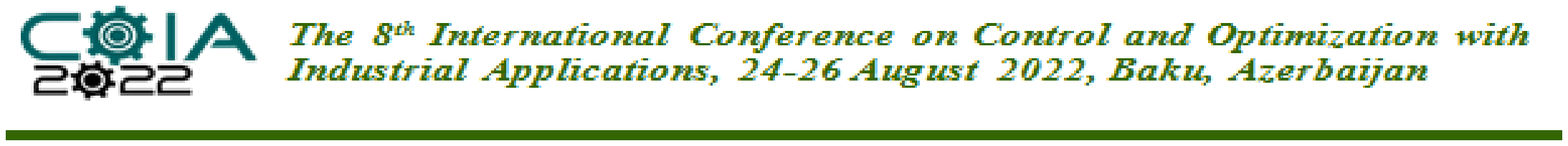}}
\end{minipage}

\bigskip
\title [Minaya Allahverdiyeva ${\rm et\, al.}$: The magnetic radii of a deuteron $R_M$ ...]
{THE MAGNETIC PROPERTIES OF A DEUTERON $R_M$ FROM THE ADS/QCD
HARD-WALL MODEL}

\author[Minaya Allahverdiyeva, Narmin Huseynova, Shahin Mamedov, Jannat Samadov] {Minaya Allahverdiyeva$^{1}$, Narmin Huseynova$^{2}$, Shahin Mamedov$^{1,2}$, Jannat Samadov$^{3}$}

 \maketitle
\begin{center}
{\scriptsize $^1$Institute of Physics, Azerbaijan National Academy of Sciences, Baku, Azerbaijan
\\ \indent $^2$Institute for Physical Problems, Baku State University, Baku, Azerbaijan
\\ \indent $^3$Shamakhy Astrophysical Observatory, Azerbaijan National Academy of Sciences, Azerbaijan

\scriptsize e-mail: minaallahverdiyeva@ymail.com, nerminh236@gmail.com, sh.mamedov62@gmail.com, jannat.samadov@gmail.com.}

\end{center}
\bigskip
\bigskip
\section{Introduction}There are two stable isotopes of hydrogen which are called protium(or hydrogen-1) and deuterium (or hydrogen-2). The deuterium also known as heavy hydrogen. The nucleus of a deuterium atom, called a deuteron and contains one proton and one neutron in the nucleus. All deuterium in the universe is thought to have been produced at the time of the Big Bang, and has endured since that time.

Since the electromagnetic (EM) properties of the deuteron can also shed light on the EM form factors of the neutron as well as on nuclear effects on the form factors, the study of the deuteron magnetic properties, especially the deuteron magnetic radius is of great interest. The deuteron is a spin–1 particle and due to current conservation and the P and C invariance of the EM interaction, it has three EM form factors in the one photon exchange (OPE) approximation, which include the charge $G_C(Q^2)$, quadrupole $G_Q(Q^2)$ and magnetic $G_M(Q^2)$ form factors and was calculated in \cite{1,2,3} at a zero temperature within soft-wall and hard-wall models AdS/QCD. In this work, we numerically calculated the deuteron magnetic radius $R_M$ in the framework of the hard-wall model of AdS/QCD and compare our results with the experimental data and soft-wall model results \cite{2}.

\bigskip
\section{Deuteron magnetic radius in the framework of the hard-wall model of ADS/QCD}
In this section we study the deuteron magnetic radii in the framework of the hard-wall model AdS/QCD. For this aim, we write an effective action, which include all interactions between the EM and deuteron fields in the bulk of the AdS space. For simplicity, we separately present all bulk Lagrangian terms, used in \cite{2,3} within the soft-wall model at $T=0$ temperature, that contribute to the $G_{1}\left( Q^{2}\right)$, $G_{2}\left( Q^{2}\right)$ and $G_{3}\left( Q^{2}\right)$ form factors of a deuteron:

1) a minimal bulk gauge action term $\textit{S}^{\left(1\right)}$:
\begin{equation} \label{eq 1}
\textit{S}^{\left(1\right)}=\int d^{4}x\ \int\limits_{0}^{z_M}dz \sqrt{g}\left[-D^M d_N^+ \left(x,z\right) D_{M} d^{N}\left(x,z\right)\right],
\end{equation}

2) a magnetic dipole bulk gauge action term $\textit{S}^{\left(2\right)}$:
\begin{equation} \label{eq 2}
\textit{S}^{\left(2\right)}=\int d^{4}x\ \int\limits_{0}^{z_M}dz \sqrt{g}\left[-ic_{2}F^{MN}\left(x,z\right) d^{+}_{M}\left(x,z\right)d_N\left(x,z\right)\right]
\end{equation}

3) a non-minimal bulk gauge action term $\textit{S}^{\left(3\right)}$:
\begin{eqnarray} \label{eq 3}
\textit{S}^{\left(3\right)}=\int d^{4}x\ \int\limits_{0}^{z_M}dz
\sqrt{g}\frac{c_{3}}{4M^2_D}\exp^{2A\left(z\right)}\partial^MF^{NK}\left(x,z\right)\left[i\partial_{K}d^{+}_{M}\left(x,z\right)d_N\left(x,z\right)-\nonumber \right.\\
-d^{+}_{M}\left(x,z\right)i\partial_K d_N\left(x,z\right)+H.C.\left.\right].
\end{eqnarray}
The common bulk action in AdS space find by summaries of these terms:
\begin{equation} \label{eq 4}
\textit{S}_{int}=\textit{S}^{\left(1\right)}+\textit{S}^{\left(2\right)}+\textit{S}^{\left(3\right)}.
\end{equation}
\begin{equation} \label{eq 5}   d^{\nu}\left(x,z\right)=\exp^{-\frac{A\left(z\right)}{2}}\sum_{n}d_{n}^{\nu}\left(x\right) \Phi_{n}\left(z\right)
\end{equation}
After carring out some calculations, performing the decomposition Eq.(5) in the Eqs.(1)-(3) and apply a Fourier transformation for the vector $V\left(x,z\right)$ and deuteron $d_{\nu}\left(x,z\right)$, $d_{\nu}^{+}\left(x,z\right)$ fields respectively, the action terms Eqs.(1)-(3) contribute to the $G_{1}(Q^{2})$, $G_{2}(Q^{2})$ and $G_{3}(Q^{2})$ form factors of the deuteron.
According to the AdS/CFT correspondence the deuteron's EM current can be found by taking a variation from the generating functional $Z=e^{iS_{int}}$ over the vector field $V_{\mu}\left(q\right)$ \cite{2,3}:
\begin{equation} \label{eq 6}
\left\langle J^{\mu}\left(p,p^{'}\right)\right\rangle=-\frac{\delta e^{iS_{int}}}{\delta V_{\mu}\left(q\right)}|_{V_{\mu}=0}.
\end{equation}
According to Eq.(6) the action terms of Eqs.(1)-(3) gives the corresponding current terms:
\begin{eqnarray} \label{eq 7}
J^{\mu\left(1\right)}(p,p^{'})=-\int dz V\left(q,z\right) \phi^{2}\left(z\right) \epsilon^{+}(p^{'}) \cdot \epsilon\left(p\right)(p+p^{'})^{\mu}=\nonumber \\
=-G_{1}\left(Q^{2}\right)\epsilon^{+}(p^{'}) \cdot \epsilon\left(p\right)(p+p^{'})^{\mu},
\end{eqnarray}
\begin{eqnarray} \label{eq 8}
J^{\mu(2)}(p,p^{'})=-c_{2} \int dz V\left(q,z\right) \phi^{2}\left(z\right) (\epsilon^{\mu}(p) \epsilon^{+}(p^{'})\cdot q-\epsilon^{+\mu}(p^{'}) \epsilon\left(p\right)\cdot q)=\nonumber \\
=-G_{2}\left(Q^{2}\right)(\epsilon^{\mu}\left(p\right) \epsilon^{+}(p^{'})\cdot q-\epsilon^{+\mu}(p^{'}) \epsilon\left(p\right)\cdot q),
\end{eqnarray}
\begin{eqnarray} \label{eq 9}
J^{\mu\left(3\right)}(p,p^{'})=\frac{c_{3}}{2M^{2}_{D}}\int dz V\left(q,z\right) \phi^{2}\left(z\right) \epsilon^{+}(p^{'})\cdot q \epsilon\left(p\right)\cdot q (p+p^{'})^{\mu}=\nonumber \\
=\frac{G_{3}\left(Q^{2}\right)}{2M^{2}_{D}}\epsilon^{+}(p^{'})\cdot q \epsilon\left(p\right)\cdot q (p+p^{'})^{\mu}.
\end{eqnarray}
The current conservation, $P$- and $C$-invariances provide three EM form factors of the deuteron, which are called the charge $G_C(Q^2)$, quadrupole $G_Q(Q^2)$ and magnetic $G_M(Q^2)$ form factors.

The EM current of a $e+d\rightarrow e+d$ electron-deuteron elastic scattering process is written in terms of $G_{i}\left(Q^{2}\right)$ form-factors:
\begin{eqnarray} \label{eq 10}
J^{\mu}(p,p^{'})=-\left(G_{1}\left(Q^{2}\right)\epsilon^{+}(p^{'})\cdot \epsilon\left(p\right)-\frac{G_{3}\left(Q^{2}\right)}{2M^2_{D}}\epsilon^{+}(p^{'})\cdot q  \epsilon\left(p\right)\cdot q \right)(p+p^{'})^{\mu}-\nonumber \\
-G_{2}\left(Q^{2}\right)\left(\epsilon^{\mu}\left(p\right) \epsilon^{+}(p^{'})\cdot q-\epsilon^{+\mu}(p^{'}) \epsilon\left(p\right)\cdot q\right).
\end{eqnarray}

According to AdS/CFT correspondence the sum of current terms Eqs.(7)-(9) is identified to the current Eq.(10).

The $G_{1}\left(Q^{2}\right)$, $G_{2}\left(Q^{2}\right)$ and $G_{3}\left(Q^{2}\right)$ are the deuteron form factors depending only upon the photon four-momentum. The explicit expressions of these form-factors within soft-wall model are found from the comparison Eqs.(7)-(9) with the current Eq.(10):
\begin{eqnarray} \label{eq 11}
G_1(Q^2)=\int dz V\left(q,z\right) \phi^{2}\left(z\right), \nonumber \\
G_2(Q^2)=c_{2} \int dz V\left(q,z\right) \phi^{2}\left(z\right), \nonumber \\
G_3(Q^2)=c_{3} \int dz V\left(q,z\right) \phi^{2}\left(z\right),
\end{eqnarray}
where $V\left(q,z\right)$ is a bulk-to-boundary propagator for the vector field and $\Phi_{n}\left(z\right)$ is a profile function of the deuteron in the framework hard-wall model AdS/QCD \cite{1}.

Thus, the momentum dependence magnetic form factors of the deuteron at zero temperature $G_M(Q^2)$ in the framework of hard-wall model AdS/QCD are related to the $G_2(Q^2)$ form-factor as~\cite{2,3}:
\begin{equation} \label{eq 12}
G_M(Q^2)=G_2(Q^2),
\end{equation}
where $\eta_{d}=\frac{Q^2}{4m^2_D}$.

From the $G_M(Q^2)$ magnetic form factor, we find magnetic radius of a deuteron in the framework of the hard-wall model as follow:
\begin{equation} \label{eq 13}
R_M=\left(-6\frac{dG_M(Q^2)}{dQ^2}|_{Q^2=0}/G_M(0)\right)^{1/2},
\end{equation}
where $G_M(0)=\frac{m_D}{m_N}\mu_D=1,714$, $m_N$ is the nucleon mass and $\mu_D=0.8574$ is a magnetic moment of a deuteron.

We calculate these derivative by use of MATHEMATICA package and then compare obtained results of the values of $R_M$ magnetic radius of a deuteron with the experimental data for this constant and with the soft-wall model results.

\bigskip
\begin{center}
	{\footnotesize  Table 1.
		
		\bigskip
		
		\begin{tabular}{|p{0.7in}|p{0.7in}|p{0.7in}|} \hline $R_M^{exp}$(fm) & {$R_M^{s.w.}$ (fm)} &
			$R_M^{h.w.}$ (fm) \\ \hline  1.90$\pm$0.14 & 2.26 & 1.48996 \\
		    \hline
		\end{tabular}}
	\end{center}
	
	\bigskip

The numerical results for the $R_M$ magnetic radius of a deuteron are presented in Table 1. Comparison of the hard-wall model results with the empirical values provided above and with the values obtained within the soft-wall model shows that the hard-wall model gives the results close to the experimental data.

\bigskip
\bigskip
{\footnotesize \noindent \textbf{Keywords:} Electromagnetic Form Factor, Infrared, Boundary, Magnetic Form Factor, Magnetic Radii.

\bigskip
\noindent \textbf{AMS Subject Classification: }83F05, 85A40.}


\bigskip
\bigskip
\begin{thebibliography}{99}
\bibitem{1} Huseynova N., Mamedov Sh., Samadov J., Deuteron electromagnetic form factors, structure functions and tensor polarization observables in the framework of the hard-wall AdS/QCD model, \emph{arxiv Preprint}, arxiv:2204.06205[hep-ph].
\bibitem{2} Gutsche T., Lyubovitskij V., Schmidt I., Deuteron electromagnetic structure functions and polarization properties in soft-wall AdS/QCD, \emph{Phys.Rev.D}, Vol.94, No.11, 2016, pp.116006.
\bibitem{3} Gutsche T., Lyubovitskij V., Schmidt I., Vega A., Nuclear physics in soft-wall AdS/QCD: Deuteron electromagnetic form factors, \emph{Phys.Rev.D}, Vol.91, No.11, 2015, pp.114001.
\end{thebibliography}
\end{document}